\documentclass[aps,prapplied,reprint,superscriptaddress]{revtex4-1}

\usepackage{mathtools}
\usepackage{graphicx}
\usepackage{physics}
\usepackage{color}
\usepackage{amssymb}
\usepackage{hyperref}
\usepackage{todonotes}

\renewcommand{\div}[1]{\vec{\nabla} \cdot #1} % for divergence
\renewcommand{\curl}[1]{\vec{\nabla} \times #1} % for curl
\let\baraccent=\= % rename builtin command \= to \baraccent

\newcommand{\Ncal}{{\cal N}}
\newcommand{\bmx}{\left( \begin{matrix}}
\newcommand{\emx}{\end{matrix} \right)}
\newcommand{\bsm}{\begin{psmallmatrix}}
\newcommand{\esm}{\end{psmallmatrix}}

\begin{document}

\title{Arbitrary control of the polarization and intensity profiles of diffraction-attenuation-resistant beams along their propagation direction}

\author{Mateus Corato-Zanarella}
\email[Corresponding author: ]{mateuscorato@gmail.com}
\affiliation{School of Electrical and Computer Engineering, University of Campinas, Campinas, SP, Brazil. Currently at Department of Electrical Engineering, Columbia University, New York, New York 10027, USA}

\author{Ahmed H. Dorrah}
\email{ahmed.dorrah@mail.utoronto.ca}
\affiliation{Edward S. Rogers Sr. Department of Electrical and Computer Engineering, University of Toronto, Toronto, Ontario M5S 3G4, Canada}

\author{Michel Zamboni-Rached}
\affiliation{School of Electrical and Computer Engineering, University of Campinas, Campinas, SP, Brazil}

\author{Mo Mojahedi}
\affiliation{Edward S. Rogers Sr. Department of Electrical and Computer Engineering, University of Toronto, Toronto, Ontario M5S 3G4, Canada}

\date{\today}
\begin{abstract}
We report on the theory and experimental generation of a class of diffraction-attenuation-resistant beams with state of polarization (SoP) and intensity that can be controlled on demand along the propagation direction. This is achieved by a suitable superposition of Bessel beams, whose parameters are systematically chosen based on closed-form analytic expressions provided by the Frozen Waves (FWs) method. Using an amplitude-only spatial light modulator, we experimentally demonstrate  three scenarios. In the first, the SoP of a horizontally polarized beam evolves to radial polarization and is then changed to vertical polarization, with the beam intensity held constant. In the second, we simultaneously control the SoP and the longitudinal intensity profile, which was chosen such that the beam's central ring can be switched-off over predefined space regions, thus generating multiple foci with different SoP and at different intensity levels along the propagation. Finally, the ability to control the SoP while overcoming attenuation inside lossy fluids is shown experimentally for the first time in the literature (to the best of our knowledge). Therefore, we envision our proposed method to be of great interest for many applications, such as optical tweezers, atom guiding, material processing, microscopy, and optical communications.
\end{abstract}

\pacs{}
\maketitle

\section{Introduction}

\par The recent advances in spatial light modulators (SLMs) and phase plates made it feasible to engineer various characteristics of light beams, thus opening new directions in light manipulation and structured light \cite{structured_light}. In most cases, non-diffracting beams have been deployed due to their interesting self-healing property and long Rayleigh range. The ability to engineer the characteristics of such beams has addressed many challenges in various fields. For instance, controlling the intensity distribution is an important feature in optical tweezers \cite{tweezer3,tweezer1} and atom guiding \cite{guiding}. Special phase distributions, such as helical phase fronts, can be designed to apply torque in micro-structures via the exchange of orbital angular momentum (OAM) \cite{OAM} and can also provide additional degrees of freedom for optical communications \cite{OAM_2}. In addition, the state of polarization (SoP) is of great interest in many areas such as material processing \cite{material_processing}, polarimetry \cite{polarimetry}, microscopy \cite{microscopy1,microscopy2} and optical communications \cite{telecom1,telecom2}.

Non-diffracting beams typically maintain their SoP with propagation. The ability to control the SoP of non-diffracting optical beams along the propagation direction is very intriguing, as such capability can potentially be deployed in various applications. For example, it can be used to spatially modulate the absorption profile of polarization-dependent optically pumped medium \cite{polarimetry}. It can also be utilized to tailor the spectrum profile of quantum emitters (which are typically polarization-dependent) over a given volume \cite{Santhosh2016}. In addition, spatially varying SoP can be used as an additional degree of freedom to control the shape and size of laser-machined structures by inducing a polarization-dependent ablation effect along the structure (in waveguide writing, for example, or to create a periodic structure) \cite{Hnatovsky:12}.

In this regard, few studies have demonstrated interesting non-diffracting beams, generated in air, with SoP that can vary continuously along the propagation direction \cite{polarization1, polarization2, polarization3, polarization4, polarization5}. The approach in Ref.~\cite{polarization1,polarization2} relied on specifying a function $\phi(r)$, whose construction is based on approximate ray description of the beam range, given that it is generated by an axicon lens. As such, achieving desired variations in the SoP within exact prescribed space intervals may require much experimentation and tailoring of $\phi(r)$, which is not systematic and may be time-consuming. The same limitation arises in the methods proposed in \cite{polarization3,polarization5}, which are based on similar procedures and do not provide a systematic recipe to design the required beam. The proposal in Ref. \cite{polarization4}, although provides a more methodical procedure, is more limited in terms of SoP control, as it only allows for trajectories along geodesics of the Poincar\'{e} sphere. Hence, special polarizations that were possible in the other methods, such as azimuthal and radial SoPs, are not addressed in this case. 

\par Additionally, the ability to control the intensity profile along the beam axis (for example, by switching the intensity of the beam center on and off or changing its intensity level with propagation) has not been shown. As a consequence, the previous techniques were only suitable for beams propagating in lossless media, that is, they did not provide means for generating optical beams that can overcome the propagation losses encountered in an absorbing medium.

In this work, we propose a systematic method to control both the SoP and the intensity profile of non-diffracting beams along the propagation direction even in the presence of medium absorption. Our proposed method relies on the Frozen Waves (FWs) theory. 

\par Frozen Waves are class of non-diffracting beams whose intensity profile can be controlled along the propagation direction. They consist of a superposition of co-propagating Bessel beams whose parameters are methodically chosen based on the desired intensity variation, as outlined in the following section. The analytical formulation of the theory provides a systematic way to design complex field structures with unusual characteristics. For instance, the FW method has been successfully applied to engineer several field characteristics, such as longitudinal intensity \cite{FW1, FW2, FWexp1, FWexp2, FW_gauss, FW_nao_zamboni_1, FW_nao_zamboni_2, compound_FW} and transverse intensity patterns \cite{compound_FW, spiral_snake}, in addition to controlling the topological charge with propagation \cite{FW_OAM} and generating spiraling and snake-like beams \cite{spiral_snake}. Furthermore, FW beams have been generated in both lossless and lossy media \cite{FW_gauss, FW_ab_1, FW_ab_2, FW_ab_3} and can be dynamically modulated over time \cite{FW_time}. 

\par The recent extension of the theory of FWs to account for vector beams with various polarization states \cite{elec_FW} opens the possibility of using these waves to also manipulate the polarization of a beam along its propagation. Here, we show how superpositions of FWs provide a systematic way to arbitrarily control the SoP and the intensity profiles of the resulting beam along its propagation, in lossless and lossy media. This is done by superposing multiple FWs with different SoPs whose contributions to the resulting beam are weighted along the propagation direction \footnote{Ultimately being switched on and off via interference if abrupt changes in SoP are desired.}, thus varying the resulting SoP. 

The validity and feasibility of the method are corroborated by three sets of experimental results obtained using an amplitude-only SLM. While our proposed method is based on analytical closed-form expressions, it is also compatible with other techniques that use FWs to manipulate other beam's characteristics, such as transverse profile and topological charge, thus proving to be a powerful and versatile tool for other advanced classes of structured light. 

\par This paper is organized as follows: in section \ref{sec. theoretical model} we summarize the FW theory and develop our proposed method for intensity and SoP control. In section \ref{example in lossy medium}, we illustrate the method with a simulated example for FW beam in a lossy medium with varying intensity and SoP profiles. In section \ref{sec: experiment}, we show and discuss the experimental results; and in section \ref{sec: conclusions} we present our conclusions.

\section{Theoretical model}
\label{sec. theoretical model}

\par A FW consists of a superposition of equal-frequency co-propagating Bessel beams (BBs) whose \textit{longitudinal intensity profile} can be arbitrarily controlled \cite{FW1, FW2, FW_gauss, FW_ab_1}. Here, the term \textit{longitudinal intensity profile} refers to the evolution of the intensity of the beam center over consecutive planes along the direction of propagation. The desired longitudinal profile, given by $|F(z)|^2$ in the interval $0\leq z \leq L$, is imprinted on a cylindrical surface of radius $\rho_0$ by using higher-order BBs or on the propagation axis with spot size $\Delta \rho_0$ by using zero-order BBs. The general form of a scalar FW in a lossy medium with refractive index $n=n_r+in_i$ is
\begin{equation}
\Psi(\rho,\phi,z,t)=e^{-i\omega t} \Ncal_{\nu}\sum^{M}_{m=-M}A_{m}J_{\nu}(h_m\rho)e^{i\nu\phi}e^{i\beta_m z}
\label{Psi}
\end{equation}

\noindent where $\Ncal_{\nu} = 1/[J_{\nu}(\cdot)]_{max}$ has the same unit of $\Psi$ and $[J_{\nu}(\cdot)]_{max}$ denotes the maximum value of the Bessel function of the first kind $J_{\nu}(\cdot)$. The longitudinal ($\beta_m=\beta_{r_m}+i\beta_{i_m}$) and transverse ($h_m$, real) wavenumbers are related by the expressions \footnote{Here, we adopted the formulation in which the beam is generated in a lossless material before penetrating the absorbing medium \cite{FW_gauss}.}
\begin{gather}
\beta_{r_m}=Q+\frac{2\pi m}{L} \\
\beta_{i_m}=\frac{\omega^2}{c^2}\frac{n_r n_i}{\beta_{r_m}} \\
h_m=\sqrt{k^2-\beta_m^2}
\end{gather}

\noindent where $k=n\omega/c=k_r+ik_i$ is the complex wavenumber and $Q$ is a constant that determines $\rho_0$ if $\nu\neq 0$ (via the first solution of $\frac{\partial}{\partial \rho}[J_\nu (h_0 \rho)]\vert_{\rho=\rho_0}=0$) or $\Delta \rho_0$ if $\nu=0$ (via $\Delta \rho_0\approx 2.4/h_0$). Superposing BBs with different wavenumbers (spatial frequencies) leads to a beating effect in the intensity profile of the envelope along the z-direction via interference. By carefully choosing the complex amplitude $A_m$ of each BB, it is possible to engineer this interference phenomenon to result in the desired intensity profile. The coefficients $A_m$ are given by \footnote{\label{coef_loss}Notice that in {Eq.~\eqref{An}} the inverse of the medium loss profile is appended to $F(z)$ in the form of $e^{\beta_{i_0}z}$, so that this augmented exponentially-growing intensity profile compensates the propagation losses.}
\begin{equation}
A_{m} = \frac{1}{L} \int^{L}_{0}F(z) e^{\beta_{i_0}z}e^{-i\frac{2\pi m}{L}z} \,\mathrm{d}z
\label{An}
\end{equation}

\noindent so that $\Psi(0,\phi,z)\approx e^{iQz}F(z)$ for $\nu=0$ or $\Psi(\rho_0,\phi,z)\approx e^{iQz}F(z)$ for $\nu\neq 0$. In the electromagnetic case, a scalar FW is assigned to the desired transverse electric field component, with the minor constraint that for azimuthal and radial polarizations the FW has azimuthal symmetry and, as a consequence, can only be of order 1. \cite{elec_FW} The longitudinal electric field component is calculated using Gauss's law (for source-free
homogeneous media, it is $\div{\vec{E}}=0$) and the magnetic field is then obtained from Faraday's law ($\vec{B}=-\frac{i}{\omega}\curl{\vec{E}}$) \footnote{Although the analysis of electromagnetic FWs in Ref. \cite{elec_FW} assumed lossless media for simplicity, all the $\vec{E}$ and $\vec{B}$ expressions presented there are valid for lossy media.}. An azimuthally polarized FW, for example, has the form $\vec{E}=E_\phi \hat{\phi}$ with 
\begin{equation}
E_\phi=e^{-i\omega t} \Ncal_1\sum^{M}_{m=-M}A_{m}J_1(h_m\rho)e^{i\beta_m z}
\label{Ephia}
\end{equation}

\noindent and has no other electric field components. \cite{elec_FW} A radially polarized FW, on the other hand, is written $\vec{E}=E_\rho \hat{\rho} + E_z  \hat{z}$ with \cite{elec_FW}
\begin{gather}
E_\rho=e^{-i\omega t} \Ncal_1 \sum^{M}_{m=-M} A_m J_1(h_m\rho)e^{i\beta_m z} \label{Erhor} \\
E_z=e^{-i\omega t} i\Ncal_1 \sum^{M}_{m=-M} \frac{h_m}{\beta_m} A_m J_0(h_m\rho)e^{i\beta_m z}
\label{Ezr}
\end{gather}

\par In addition, a linearly polarized FW of order $\nu$ in the $\hat{x}$ or $ \hat{y}$ direction has the form \cite{elec_FW}
\begin{gather}
\vec{E} = E_\perp \bsm  \hat{x} \\  \hat{y} \esm + E_z  \hat{z} \\
E_\perp = e^{-i\omega t}\Ncal_{\nu}\sum^{M}_{m=-M}A_{m}J_{\nu}(h_m\rho)e^{i\nu\phi}e^{i\beta_m z}
\label{E_FW} \\
\begin{align}
\nonumber E_z &= e^{-i\omega t}\Ncal_\nu \sum_{m=-M}^{M} A_m e^{i\beta_m z} e^{i\nu \phi} \\
&\times \left[\bsm \sin{\phi} \\ -\cos{\phi} \esm \frac{\nu}{\rho\beta_m}J_\nu + \frac{(J_{\nu-1}-J_{\nu+1})}{2} i \bsm \cos{\phi} \\ \sin{\phi} \esm \frac{h_m}{\beta_m}\right] 
\label{Ezlx}
\end{align}
\end{gather}

\noindent where the brackets notation is used to distinguish the terms corresponding to each direction and $J_\nu \equiv J_\nu(h_m\rho)$. Any polarization state in the Poincar\'e sphere can be obtained by combining orthogonal linearly polarized FWs with suitable complex amplitudes $a_x$ and $a_y$, that is, $\vec{E}=(a_x \hat{x} + a_y \hat{y})E_\perp + E_z \hat{z}$ where $E_z$ is a combination of results of the type \eqref{Ezlx}. 

\par The key to understand how FWs can be used to control the polarization of a beam along its propagation direction is to notice that the SoP and the longitudinal intensity profile of a FW are decoupled. This means that any desired polarization state is compatible with any arbitrary longitudinal intensity profile. Additionally, the transverse characteristics of each FW can also be tailored by suitable choices of its order \footnote{This degree of freedom is not available, however, in the case of azimuthal and radial polarizations, for which the order is fixed at 1 \cite{elec_FW}.} and spot size (for zero order) or radius (for higher orders), via the value of $Q$. As a consequence, the following procedure may be used to generate a beam with desired polarization and intensity profiles along the propagation direction:

\begin{enumerate}
	
	\item Define the intervals in which the beam will posses a particular polarization state;
	
	\item Define the desired longitudinal intensity profile within each interval;
	
	\item For each interval, assign a FW with the necessary phase, polarization and intensity patterns, making its intensity negligible outside of it. \label{step for choosing FWs}
	
	\item For the experimental generation, decompose each FW into its $x$- and $y$-polarized components. \label{step4}
	
\end{enumerate}

\par In step \ref{step for choosing FWs}, the transverse characteristics of the FWs are determined by choosing their spot sizes or radii and their orders, which define their topological charges. A more advanced control can be achieved by combining different FWs with the same polarization but different longitudinal intensity profiles, spot sizes or radii and orders, as presented in other works \cite{compound_FW, spiral_snake, FW_OAM}. 
The step \ref{step4} is necessary for generating the resulting field using an SLM, as further explained in Sec. \ref{sec: experiment}.

\par This approach allows both abrupt and continuous longitudinal variations in the SoP. If the intensity profiles of FWs with different polarizations do not overlap along the longitudinal direction, the result will be a pattern with abrupt changes. On the other hand, the change will be continuous wherever there is an overlap. Since the complex amplitude of each FW is determined by $F(z)$, we have total freedom in composing these variations. As an example, we may combine two linearly polarized FWs with orthogonal electric field components such that, by adjusting $F(z)$ of each, the resulting SoP may undergo any trajectory in the Poincar\'e sphere as the propagation distance increases.

\section{Example in a lossy medium and discussions}
\label{example in lossy medium}

\par To illustrate the outlined procedure, we present an example with abrupt polarization and intensity variations in an absorbing medium, which is assumed to have a complex refractive index of $n=n_r+in_i=2+i7.5\times 10^{-7}$ at the wavelength $\lambda=532\,\text{nm}$. The total interval in which the field is to be modeled ranges from $z=0$ to $z=L=17\,\text{cm}$ and all the FWs used have order $\nu=1$, with $M=30$ and $Q=0.9999k_r$, resulting in a constant radius of $\rho_0\approx 6.56\,\mu\text{m}$.

\par Since all the FWs have order $\nu=1$, we can introduce a compact notation for the resulting electric field:
\begin{subequations}
	\begin{gather}
	\vec{E}(\rho,\phi,z,t) =  E_x \hat{x} + E_y \hat{y} \label{Eq1} \\
	E_{u}(\rho,\phi,z,t)|_{u=x,y} = e^{-i\omega t}\, \sum_{p=1}^{N} G_{u p}(\phi) \sum_{m=-M}^{M} A_{p m} J_{1}(h_{p m} \rho) e^{i\beta_{p m} z}
	\end{gather}
	\label{compact_notation}
\end{subequations}

\noindent where $N$ is the number of superposed FWs and $G_{up}(\phi)$ is a morphological function with the azimuthal dependence, which is $e^{i\phi}$ for linear polarization (due to the structure of linearly-polarized FWs) and $\cos(\phi)$ or $\sin(\phi)$ when decomposing other states of polarization, such as radial and azimuthal, into their $x$- and $y$-polarized components. Additionally, each FW has a function $F_p(z)$ that defines its longitudinal intensity profile. Notice that this notation is just another representation that is used here to write the superposition of FWs in a compact way and is in total agreement with the theory in Sec. \ref{sec. theoretical model}.

\par In this example, the desired characteristics of the total field are:

\begin{enumerate}
	\item For $0\leq z < 5\,\text{cm}$, the polarization is linear in the x-direction and the intensity is constant and equal to $1\,\text{arb. units}$.
	
	\item For $5\,\text{cm} \leq z < 10\,\text{cm}$, the polarization is azimuthal and the intensity is constant and equal to $2\,\text{arb. units}$.
	
	\item For $10\,\text{cm} \leq z < 15\,\text{cm}$, the polarization is radial and the intensity is constant and equal to $3\,\text{arb. units}$.
	
	\item For $15\,\text{cm} \leq z < 17\,\text{cm}$, the field is negligible over the cylindrical surface.
	
\end{enumerate}

\par Therefore, the beam should be composed of three FWs: the first with linear polarization in the $x$-direction and $F_1(z)=1$ within $0\leq z < 5\,\text{cm}$; the second with azimuthal polarization and $F_2(z)=\sqrt{2}$ within $5\,\text{cm}\leq z < 10\,\text{cm}$; and the third with radial polarization and $F_3(z)=\sqrt{3}$ within $10\,\text{cm} \leq z < 15\,\text{cm}$. Outside the mentioned intervals, the FWs have $F_p (z)=0$, which means that their inner rings are designed to switch-off over this region. In the notation of Eq. \eqref{compact_notation}, we have $G_{x1}=e^{i\phi}$, $G_{y1}=0$, $G_{x2}=-\sin(\phi)$, $G_{y2}=\cos(\phi)$, $G_{x3}=\cos(\phi)$, $G_{y3}=\sin(\phi)$ and
\begin{align}
\hspace*{-6mm}\begin{cases}
F_1(z)=1\,\text{,} \ \ F_2(z)=0\,\text{,}  \! \quad \ \ F_3(z)=0 \quad \, \ \text{for } 0\,\text{cm}\leq z < 5\,\text{cm} \\
F_1(z)=0\,\text{,} \ \ F_2(z)=\sqrt{2}\,\text{,} \ \ F_3(z)=0 \quad \, \ \text{for } 5\,\text{cm}\leq z < 10\,\text{cm} \\
F_1(z)=0\,\text{,} \ \ F_2(z)=0\,\text{,} \! \quad \ \ F_3(z)=\sqrt{3} \ \ \text{for } 10\,\text{cm}\leq z < 15\,\text{cm} \\
F_1(z)=0\,\text{,} \ \ F_2(z)=0\,\text{,} \! \quad \ \ F_3(z)=0 \ \quad \, \text{elsewhere}
\end{cases}
\label{example_parameters}
\end{align}

\par
\begin{figure}[h]
	\centering
	\includegraphics[width=\columnwidth]{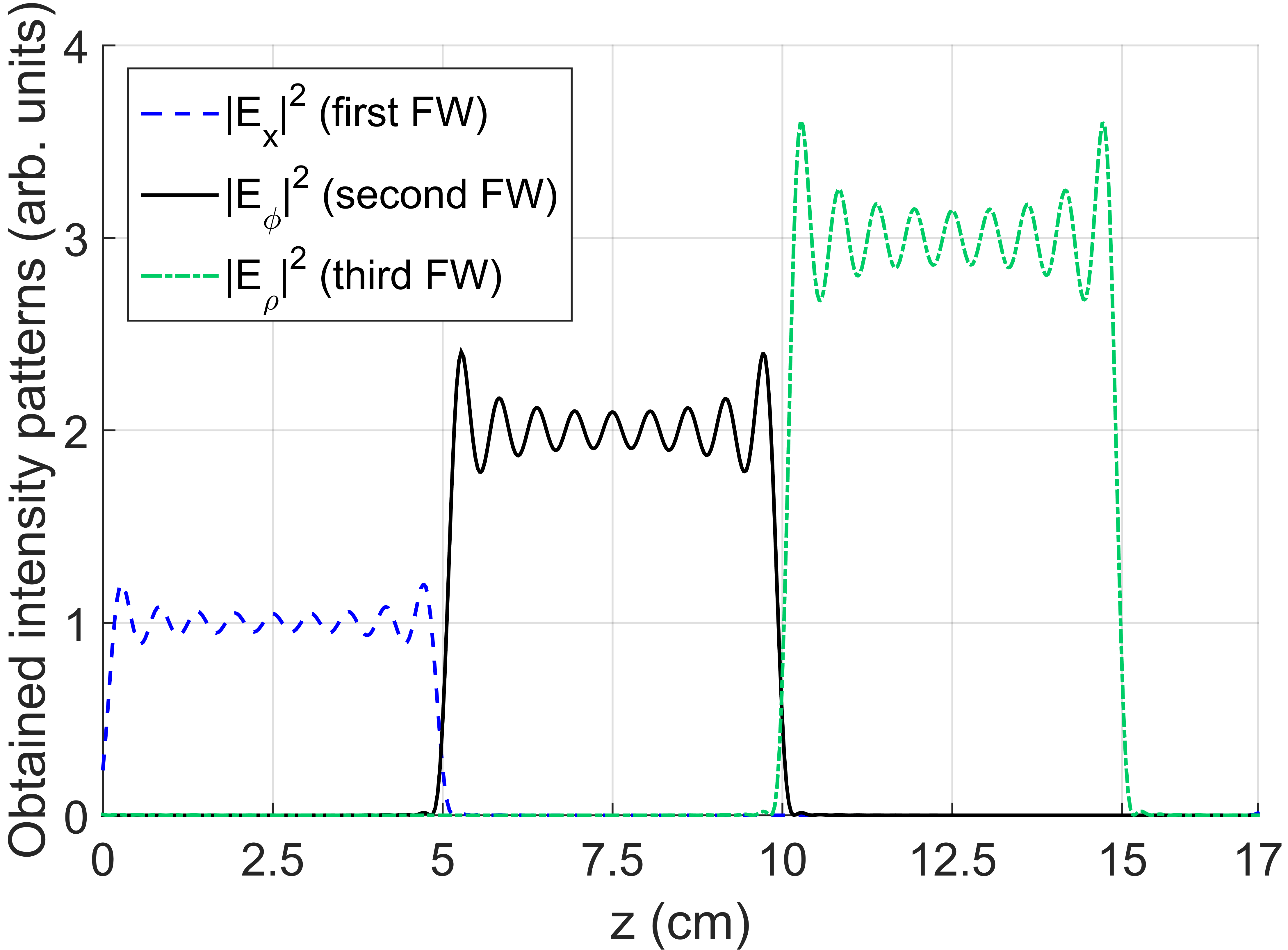}
	\caption{Obtained intensity patterns of the FWs in each interval for the example given. We see that the contribution of only one FW dominates inside each region.}
	\label{intensity_pattern_paper}
\end{figure}

 Fig. \ref{intensity_pattern_paper} depicts the obtained intensity patterns for each of the superposed FWs and shows that the desired abrupt changes in the polarization occur exactly at the prescribed longitudinal positions. Since the curves overlap a little, the actual transitions are continuous. These brief overlaps could be reduced by further increasing the value of $M$ of each FW.

\par
\begin{figure}[h]
	\centering
	\includegraphics[width=\columnwidth]{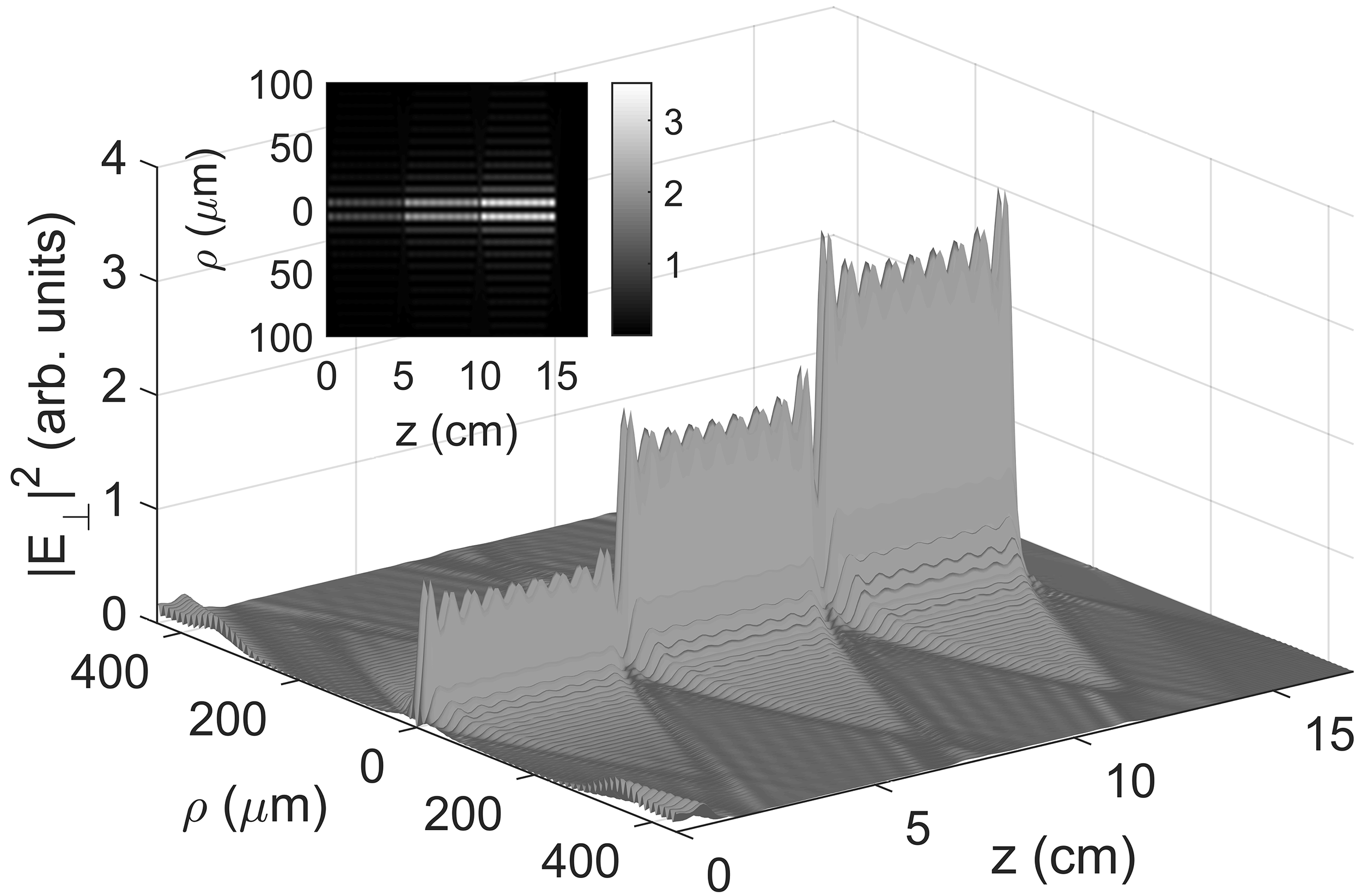}
	\caption{Intensity of the transverse electric field, $|\vec{E}_\perp|^2$, of the example given, which is in agreement with the desired abrupt intensity changes.}
	\label{plot2D}
\end{figure}

 The resulting 3D intensity profile of the transverse electric field is shown in Fig. \ref{plot2D} and is in agreement with Fig. \ref{intensity_pattern_paper}. Notice that $15\,\text{cm}$ is more than twice the penetration depth \cite{FW_gauss} $1/(2k_i)\approx 6.71\,\text{cm}$ of an ordinary beam propagating in the same medium. Despite that, not only does the obtained pattern overcome the propagation losses, but it also follows the desired intensity profile, thus confirming the attenuation-resistant feature of the FWs \cite{FW_ab_3}. Moreover, note that the field has a good transverse localization with progressively less intense lateral rings, which is also characteristic of a FW.

\par
\begin{figure*}[htbp]
	\centering
	\includegraphics[width=0.9\textwidth]{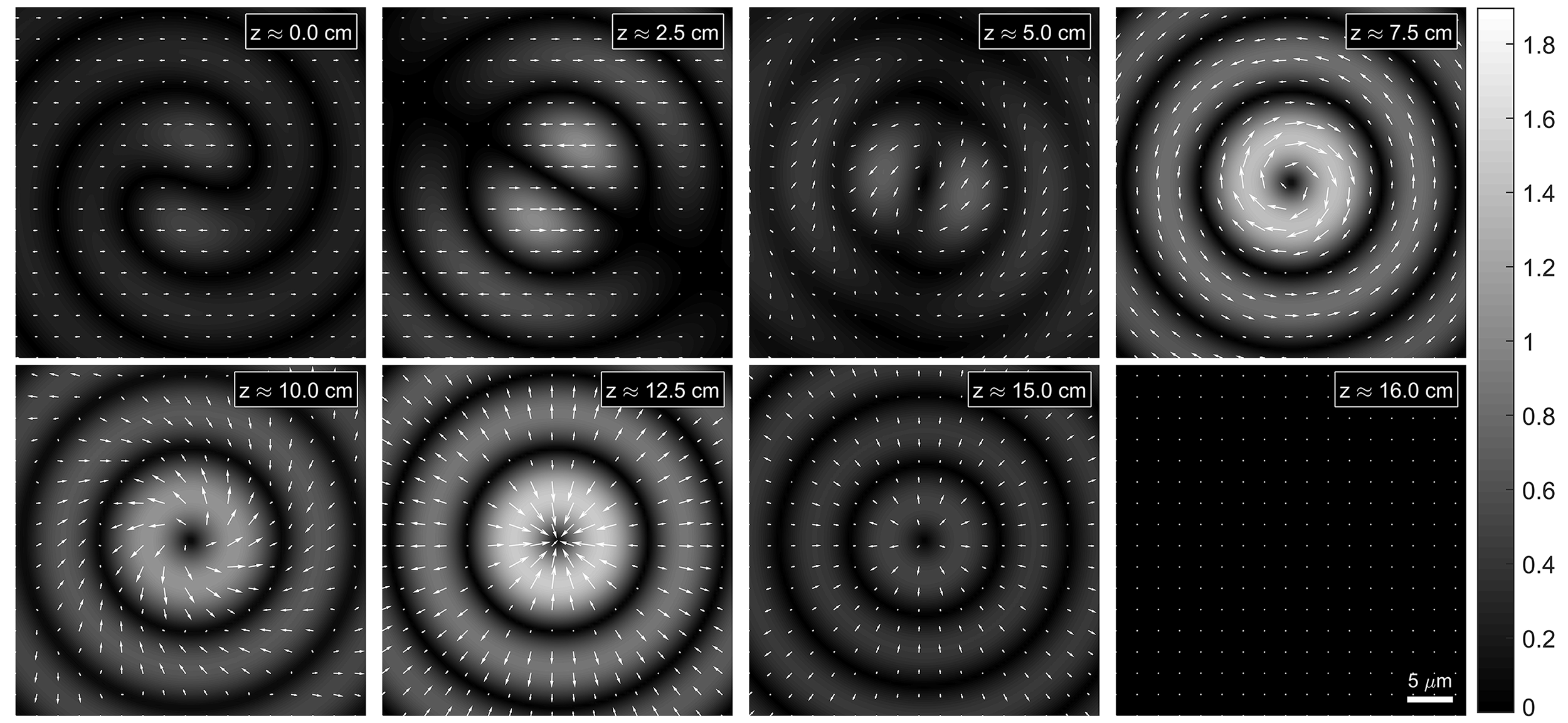}
	\caption{Polarization and amplitude of the real transverse electric field ($|\Re[\vec{E}_\perp]|$) as $z$ increases for the given example, with the time is fixed at $t=0$. Inside $0\,\text{cm}\leq z < 5\,\text{cm}$, the field has linear polarization in the $x$-direction and an amplitude of $1$ arb. units. Then, at $z=5\,\text{cm}$, the polarization changes to azimuthal and the amplitude becomes $\sqrt{2}$ arb. units (corresponding to an intensity of $2$ arb. units). These characteristics are kept until $z=10\,\text{cm}$, where the SoP evolves to radial polarization and the amplitude is increased to $\sqrt{3}$ arb. units (corresponding to an intensity of $3$ arb. units). The field is then switched-off at $z=15\,\text{cm}$ and the amplitude of the inner ring becomes negligible afterwards.}
	\label{frames}
\end{figure*}

 Fig. \ref{frames} shows how the SoP and the amplitude of the real transverse electric field ($|\Re[\vec{E}_\perp]|$) change when $z$ increases and the time is fixed at $t=0$. Indeed, we see the abrupt variations in both polarization and amplitude predicted by Fig. \ref{intensity_pattern_paper}, in agreement with the desired characteristics. Such abrupt transitions are made possible by including large number of BBs in the superposition ($M=30$), some of which have higher spatial frequencies to construct sharp variations in the longitudinal intensity profile.

\par Notice that Fig. \ref{frames} does not depict the intensity pattern of the transverse field ($|\vec{E}_\perp|^2$), but actually the amplitude of the real transverse field ($|\Re[\vec{E}_\perp]|^2$). Therefore, in the case of the linear polarization, the phase of the form $e^{i(\phi+\beta z)}$ of the complex field becomes $\cos(\phi+\beta z)$ after taking the real part, thus generating a pattern with two petals. However, the intensity pattern of the transverse field is ring-shaped, as the ones of azimuthal and radial polarizations  \footnote{This is also shown in the experimental results of Sec. \ref{sec: experiment}.}.

\par By merely observing the transitions in the state of the central ring in Fig. \ref{frames}, one might argue that the energy and the momenta (linear and angular) of the beam are not conserved, but in fact no physical law is violated. The unusual evolution of the beam is due to a continuous spatial redistribution of these quantities due to the interference of BBs, caused by the judicious choice of their complex amplitudes via Eq.~\eqref{An}. While energy and momenta densities may vary during propagation and give the impression that they are not conserved locally, the total quantities remain the same across each transverse plane. For example, when the beam's intensity decreases near the axis after $z=15\,\text{cm}$, it increases along the radial direction of the field, diffusing the energy and the momenta far from the axis. This is exactly the essence behind the theory of FWs: engineering the interference of BBs to exchange energy and momenta between the center of the beam and its lateral structure at will.

\section{Experimental generation and results}
\label{sec: experiment}

\par
\begin{figure*}[htbp]
	\centering
	\includegraphics[width=.825\textwidth]{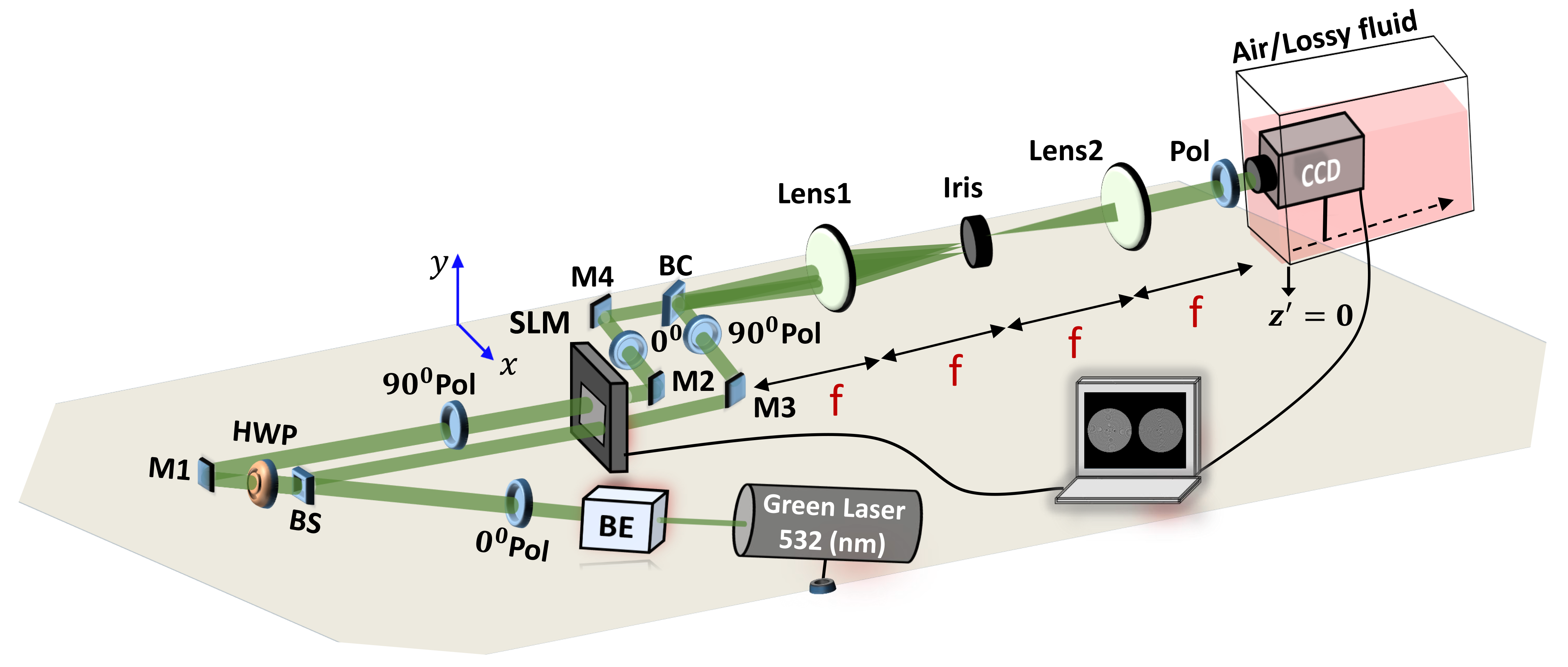}
	\caption{Experimental setup used to demonstrate longitudinal control of the polarization state and intensity. An amplitude-only transmissive spatial light modulator is addressed by a computer generated hologram that is designed at 532 nm wavelength. Two collimated beams with linear and orthogonal polarization states are incident on the SLM screen, where each beam is encoded on one half of it. The output patterns are combined and then filtered using a 4-$f$ system. The combined beam is then detected by a CCD camera along the beam axis. For the case of beam propagation inside a lossy fluid, the CCD camera is immersed inside the container while covered by a waterproof case. A polarizer is used to analyze the polarization state at the plane of the CCD camera. Here, BE is the beam expander, BS refers to the beam splitter, BC is the beam combiner, M refers to the mirrors, HWP is the half wave plate, Pol refers to the polarizers, and SLM is the spatial light modulator.}
	\label{Fig1}
\end{figure*}

 We experimentally demonstrate three different beam patterns with control over SoP along the propagation direction. The first two patterns were generated in air, whereas the third pattern was generated inside a lossy fluid. The experimental setup is depicted in Fig.~\ref{Fig1}. 

\par The desired transverse electric field at the initial plane $\vec{E}_\perp(\rho,\phi,z=0,t)$ is decomposed into two components with linear and orthogonal SoP along $x$ and $y$ directions. Each component, $E_x(\rho,\phi,z=0,t)$ and $E_y(\rho,\phi,z=0,t)$, is calculated and transformed into a 2D computer generated hologram (CGH) that is mapped onto one part of the SLM. Here, we used the HOLOEYE LC2012 SLM, which has a twisted nematic liquid crystal transmissive screen that operates in the amplitude-only mode. The SLM screen is divided into two adjacent sections, each being addressed independently by $E_x$ and $E_y$. A 532 nm collimated laser beam with linear polarization along the vertical axis $y$ is split into two beams via the beam splitter `BS1'. The first beam, which is now $y$-polarized, passes through the right side of the SLM. The second beam passes through a half wave plate (HWP) that is rotated at an angle of 45$^\circ$ with respect to the vertical axis. This rotates the polarization state of the beam by an angle of 90$^\circ$ (i.e. rotating it to the orthogonal direction $x$). Such beam, now $x$-polarized, then passes through the left side of the SLM. Due to the twisted nematic nature of the SLM, the output patterns encoded from the CGHs on the right and left sides of the SLM then pass through polarizers oriented at $90^\circ$ and $0^\circ$ with respect to the vertical axis, respectively. This helps to clean the encoded patterns from unwanted polarizations and generate high contrast images.

It should be noted that we have used an amplitude mask to express the complex transmission function $E_u(\rho,\phi,z=0,t)|_{u=x,y}$. The hologram equation can be mathematically expressed as
\begin{equation}
\label{Eq5}
H(x,y) = \frac{1}{2}\lbrace\beta(x,y)+\alpha(x,y)\cos[\Phi(x,y)-2\pi(u_0x + v_0y)]\rbrace
\end{equation} 

\noindent where $\alpha(x,y)$ and $\Phi(x,y)$ are the amplitude and phase of $E_u(\rho,\phi,z=0,t)$, respectively. A bias function $\beta(x,y)$ is chosen as a soft envelope for the amplitude $\alpha(x,y)$ and is given by $\beta(x,y) = [1+\alpha(x,y)^2]/2$ \cite{FWexp1,FWexp2,FW_OAM,FW_ab_3}. 

\par The two output beams, with orthogonal polarization states, are then combined via the beam splitter 'BS2', resulting in the vector waveform $\vec{E}(\rho,\phi,z,t)$. The resulting beam is then imaged and filtered using a 4-$f$ system (with $f=20$ cm) and an iris. For efficient filtering, we superpose a plane wave $\exp[2\pi i(u_0x + v_0y)]$ on the computer generated hologram. This shifts the encoded pattern off-axis to the spatial frequencies ($u_0,v_0$) in the Fourier plane; thus making it easier to filter out the shifted pattern from the undesired on-axis noise by using an iris. The parameters $u_0$ and $v_0$ were set to $3/(16\Delta{x})$, where $\Delta{x}$ is the SLM pixel pitch (36$\,\mu$m). The filtered pattern is then imaged back at the focus of `Lens2' which we refer to as the $z'=0$ plane. We note that this plane maps to an actual propagation distance of $20\,$cm along the beam axis, as this is the approximate distance it propagates after the SLM and before it is imaged by the 4-$f$ system. The evolution of the generated waveform is then recorded using a CCD camera along the beam axis either in air or inside a lossy fluid. At each plane of detection, the polarization state of the measured beam is analyzed by a polarizer oriented at 0$^\circ$, 90$^\circ$, 45$^\circ$, and 135$^\circ$, with respect to the vertical axis $y$, to verify the change in the SoP with propagation. 

\par We have experimentally generated three different beam patterns with $M=5$. Accordingly, each of the $x$- and $y$-polarized components $E_{x}$ and $E_{y}$ of the FWs consists of $2M+1=11$ equal-frequency Bessel beams of order $\nu=1$ with equally spaced longitudinal wavenumbers. We assigned values of $0.999995k_r$ and $0.9999978k_r$ to the parameter $Q$ for the FWs generated in air and in the lossy fluid, respectively. By doing so, the generated Bessel beams possess transverse wavenumbers that are compatible with the spatial bandwidth of the available SLM. In all cases, we set $L$ = $1\,$m. Here, we will use the same compact notation adopted in Sec. \ref{example in lossy medium}.

\par In the first pattern, generated in air, we show control over the SoP while the longitudinal intensity profile is kept constant along the propagation by superposing three different FWs. Their morphological functions are $G_{x1}(\phi)=e^{i\phi}$, $G_{y1}(\phi)=0$, $G_{x2}(\phi)=\cos(\phi)$, $G_{y2}(\phi)=\sin(\phi)$, $G_{x3}(\phi)=0$, $G_{y3}(\phi)=e^{i\phi}$ and their functions $F_p(z)$ are
\begin{align}
\hspace*{-1.5mm}\begin{cases}
F_1(z)=1\,\text{,}  \ \ F_2(z)=0\,\text{,}  \ \ F_3(z)=0 \ \ \ \underset{\text{\normalsize(x-polarization)}}{\text{for } 0\,\text{cm}\leq z^\prime < 10\,\text{cm}} \\
F_1(z)=0\,\text{,}  \ \ F_2(z)=1\,\text{,} \ \ F_3(z)=0 \ \ \ \underset{\text{\normalsize(radial polarization)}}{\text{for } 10\,\text{cm}\leq z^\prime < 25\,\text{cm}} \\
F_1(z)=0\,\text{,}  \ \ F_2(z)=0\,\text{,}  \ \ F_3(z)=1 \ \ \ \underset{\text{\normalsize(y-polarization)}}{\text{for } 25\,\text{cm}\leq z^\prime < 35\,\text{cm}}
\\
F_1(z)=0\,\text{,}  \ \ F_2(z)=0\,\text{,}  \ \ F_3(z)=0 \ \ \ \text{elsewhere}
\end{cases}
\label{Equation2}
\end{align}

\noindent where the resulting polarization is indicated under each interval.

\begin{figure*}[htbp]
	\centering
	\includegraphics[width=0.825\textwidth]{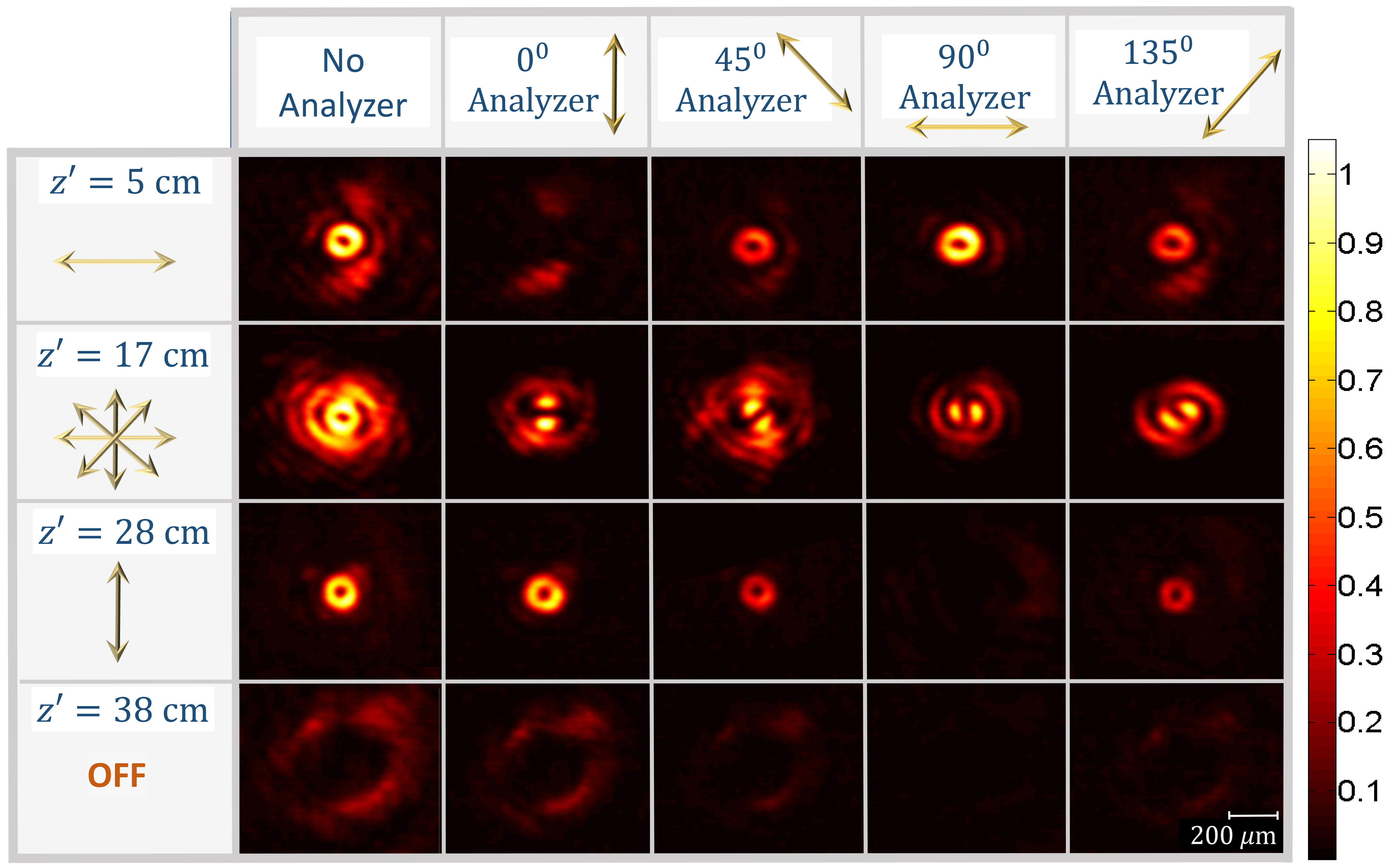}
	\caption{Normalized intensity profiles of the measured beam for the first example (generated in air). The top row indicates the analyzer angle used before the CCD camera to verify the change in SoP with propagation and the leftmost column indicates the desired SoP. The angles are measured with respect to the vertical axis $y$. The beam is linearly polarized in the $x$-direction over the range of $0\,\text{cm}\leq z^\prime < 10\,\text{cm}$, before the SoP evolves to radial polarization, which is observed at $z^\prime=17\,$cm. Then, the SoP is changed to linear polarization in the $y$-direction, as seen at $z^\prime=28\,$cm. Finally, the beam center is switched-off, as shown at $z'=38\,$cm. In all the SoPs, the longitudinal intensity profile is kept constant along the beam axis, as seen in the first column (``No Analyzer'').}
	\label{Fig2}
\end{figure*}

\par According to these parameters, the resulting beam has the following properties: it is $x$-polarized over the range $0\,\text{cm}\leq z^\prime < 10\,\text{cm}$, since only the $E_x$ of the first FW is significant near the beam center, with all the contributions to $E_y$ and $E_x$ of the other FWs being stored in the outer rings, away from the axis; then, the beam becomes radially polarized over the range $10\,\text{cm}\leq z^\prime < 25\,\text{cm}$, having equal in-phase contributions from $E_x$ and $E_y$  \footnote{We note that a phase bias (retardation) has been added in the path of $E_y$ to compensate for any phase difference with respect to $E_x$, thus ensuring radial polarization. Such phase bias is fixed for all measurements along the beam axis.}; afterwards, the SoP becomes $y$-polarized over the range $25\,\text{cm}\leq z^\prime < 35\,\text{cm}$, which is done by switching off the contributions of $E_x$; finally, all the FWs are switched-off by assigning them $F(z)=0$ after the previous interval, so that the inner ring vanishes because the energy is stored away from the center of the beam into the outer rings via an inverse self-healing process. The evolution of the transverse profile of the resulting beam recorded with the CCD camera at different propagation distances is shown in Fig.~\ref{Fig2}. 

\par The arrows in the left column indicate the predefined SoP of the beam, whereas the arrows in the top row depict the analyzer angle. As discussed, the beam exhibits linear SoP in the $x$-direction at $z^\prime=5\,\text{cm}$. This is evident by looking at the transverse intensity profile, which is at maximum level when the analyzer is oriented along the $x$-direction ($90^\circ$ with respect to the vertical axis $y$). In contrast, the recorded intensity becomes negligible when the analyzer is in the orthogonal direction (along the $y$-direction) and reaches approximately half the intensity level in between. Afterwards, at the distance of $17\,\text{cm}$, the SoP is altered and the beam becomes radially polarized. This is verified by looking at the intensity profile, which now has the form of two petals recording maximum intensity level along the analyzer axis and zero intensity in the orthogonal direction. Then, at $z^\prime=28\,\text{cm}$, the SoP is $y$-polarized, as shown by the results in the third row, which represent the complementary scenario of the first row ($z^\prime=5\,\text{cm}$). Finally, the beam center is switched-off, as observed at $z^\prime=38\,\text{cm}$. Notice that the peak intensity is kept constant during propagation, as seen in the ``No Analyzer'' column. 

\par From Fig.~\ref{Fig2}, it can also be observed that the intensity of the outer rings changes with propagation. This is a typical behavior that results from interfering multiple Bessel beams with different (transverse) wavenumbers. The outer rings observed at $z^\prime= 5$ and $17$ cm get focused throughout propagation and are responsible for constructing the $y$-polarized inner ring at $z^\prime = 28$ cm. When the inner ring is designed to switch off, for example at the plane $z^\prime=38$ cm, the energy in the inner ring is redistributed to the outer rings again. At this same position, the outer rings that were responsible for the $x$-polarized part of the beam cannot be observed, as they already interfered previously and are spread outside of the observation window. 

\begin{figure*}[htbp]
	\centering
	\includegraphics[width=.825\textwidth]{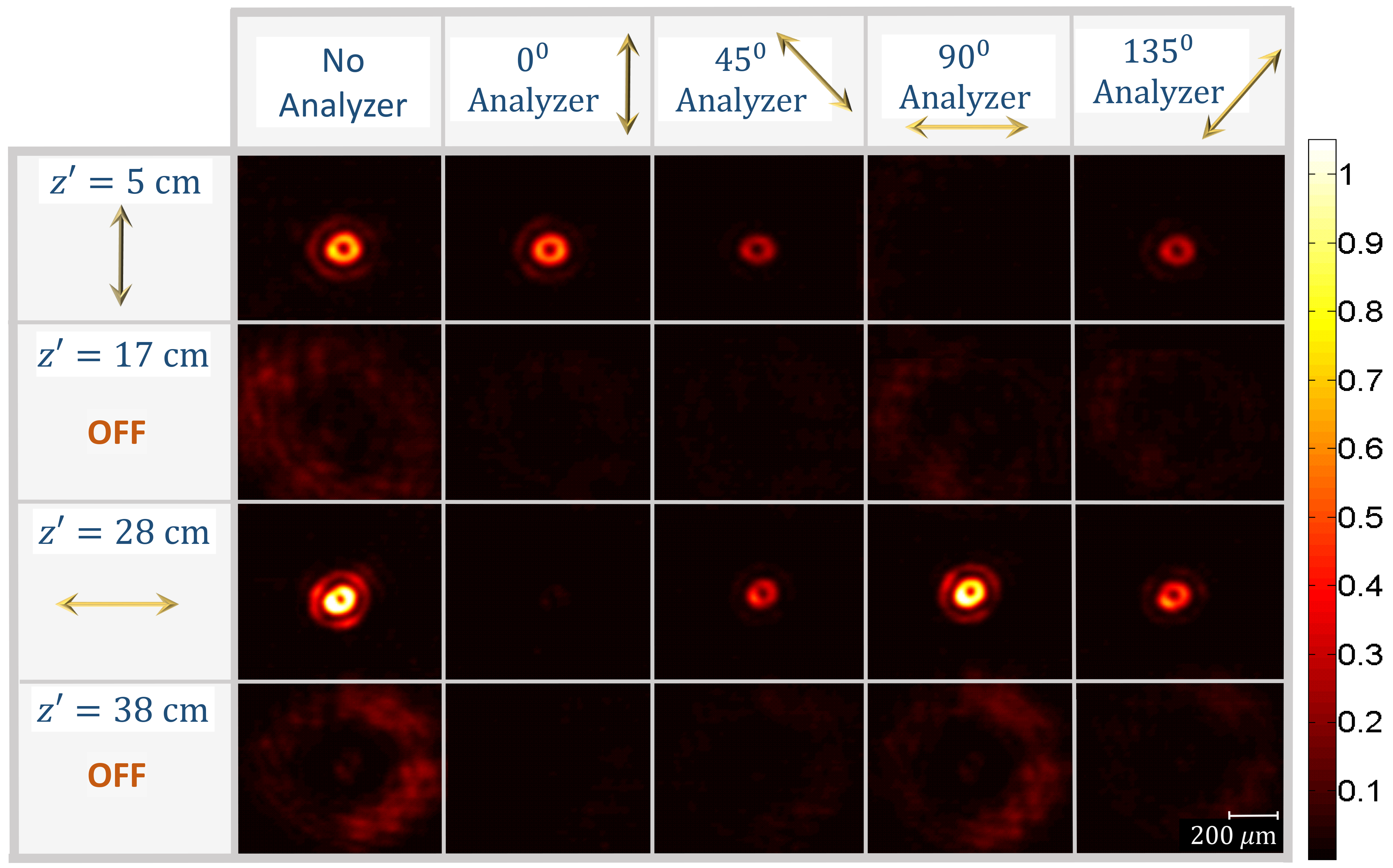}
	\caption{Normalized intensity profiles of the measured beam for the second example (generated in air). The top row indicates the analyzer angle used before the CCD camera to verify the change in SoP with propagation and the leftmost column indicates the expected SoP. The angles are measured with respect to the vertical axis $y$. The beam is linearly polarized in the $y$-direction over the range $0\,\text{cm}\leq z^\prime < 10\,\text{cm}$ before it is switched-off for the interval $10\,\text{cm}\leq z^\prime < 25\,\text{cm}$. Then, it is made to retain a linear polarization in the $x$-direction with double the original intensity level over the range $25\,\text{cm}\leq z^\prime < 35\,\text{cm}$, after which the beam is switched-off again.}
	\label{Fig3}
\end{figure*}

\par This example demonstrates that our approach of controlling the intensity of FWs with propagation makes it possible to change the SoP of the beam in a systematic manner as it propagates, while keeping the same intensity level. Indeed, this approach is very flexible and allows the generation of numerous other interesting patterns along the longitudinal direction. For instance, in the second pattern generated in air, we show the possibility of controlling both the SoP and longitudinal intensity profile along the beam axis. In this scenario, $E_x$ and $E_y$ are designed to exhibit different intensity levels by using two linearly-polarized FWs with orthogonal polarizations. The morphological functions are $G_{x1}=0$, $G_{y1}=e^{i\phi}$, $G_{x2}=e^{i\phi}$, $G_{y2}=0$ and their functions $F_p(z)$ are
\begin{align}
\begin{cases}
F_1(z)=1\,\text{,}  &F_2(z)=0 \qquad \! \underset{\text{\normalsize(y-polarization)}}{\text{for } 0\,\text{cm}\leq z^\prime < 10\,\text{cm}}
\\
F_1(z)=0\,\text{,}  &F_2(z)=0 \qquad \! \text{for } 10\,\text{cm}\leq z^\prime < 25\,\text{cm} \\
F_1(z)=0\,\text{,}  &F_2(z)=\sqrt{2} \quad \underset{\text{\normalsize(x-polarization)}}{\text{for } 25\,\text{cm}\leq z^\prime < 35\,\text{cm}}
\\
F_1(z)=0\,\text{,}  &F_2(z)=0 \qquad \! \text{elsewhere}
\end{cases}
\label{Equation3}
\end{align}

\noindent where the resulting polarization is indicated below each interval.

\par Accordingly, the generated beam is $y$-polarized over the range $0\,\text{cm}\leq z^\prime < 10\,\text{cm}$, since it mainly has the contribution of $E_y$ in the beam center. Then, in the interval $10\,\text{cm}\leq z^\prime < 25\,\text{cm}$, the beam is switched-off by assigning $F_p(z)=0$ to both $E_x$ and $E_y$. Afterwards, it becomes $x$-polarized over the range $25\,\text{cm}\leq z^\prime < 35\,\text{cm}$. This is done simultaneously to an intensity level increase, such that it becomes double the value in the first interval. Finally, the beam center is made to switch-off by assigning a value of 0 for $F_p(z)$ of both components $E_x$ and $E_y$. The evolution of the transverse profile of the resulting beam recorded with the CCD camera at different propagation distances is shown in Fig.~\ref{Fig3}. 

\par As discussed, the beam exhibits linear SoP in the $y$-direction at $z^\prime=5\,$cm, which is evident from the transverse intensity profile that records maximum level when the analyzer is oriented along the $y$-direction and negligible intensity along the orthogonal direction. The beam center is then switched-off, as shown at $z^\prime=17\,$cm, before it is switched-on again at twice the intensity level with an SoP in the $x$-direction, as shown at $z^\prime=28\,$cm.  Finally, the beam center is switched-off, as observed at $z^\prime=38\,$cm. 

\par In order to demonstrate the attenuation-resistance property of FWs, we present the third and last example, in which the beam changes its SoP from $x$-polarized to $y$-polarized while keeping a constant intensity level throughout propagation inside a lossy fluid, thus overcoming absorption. The fluid was prepared by adding a few drops of propylene glycol, citric acid and sodium benzoate to $4\,\text{L}$ of water. The resulting fluid has a complex index of refraction given by $n=1.4 + i 0.5\times10^{-6}$ at 532 nm. 

\begin{figure*}[htbp]
	\centering
	\includegraphics[width=.85\textwidth]{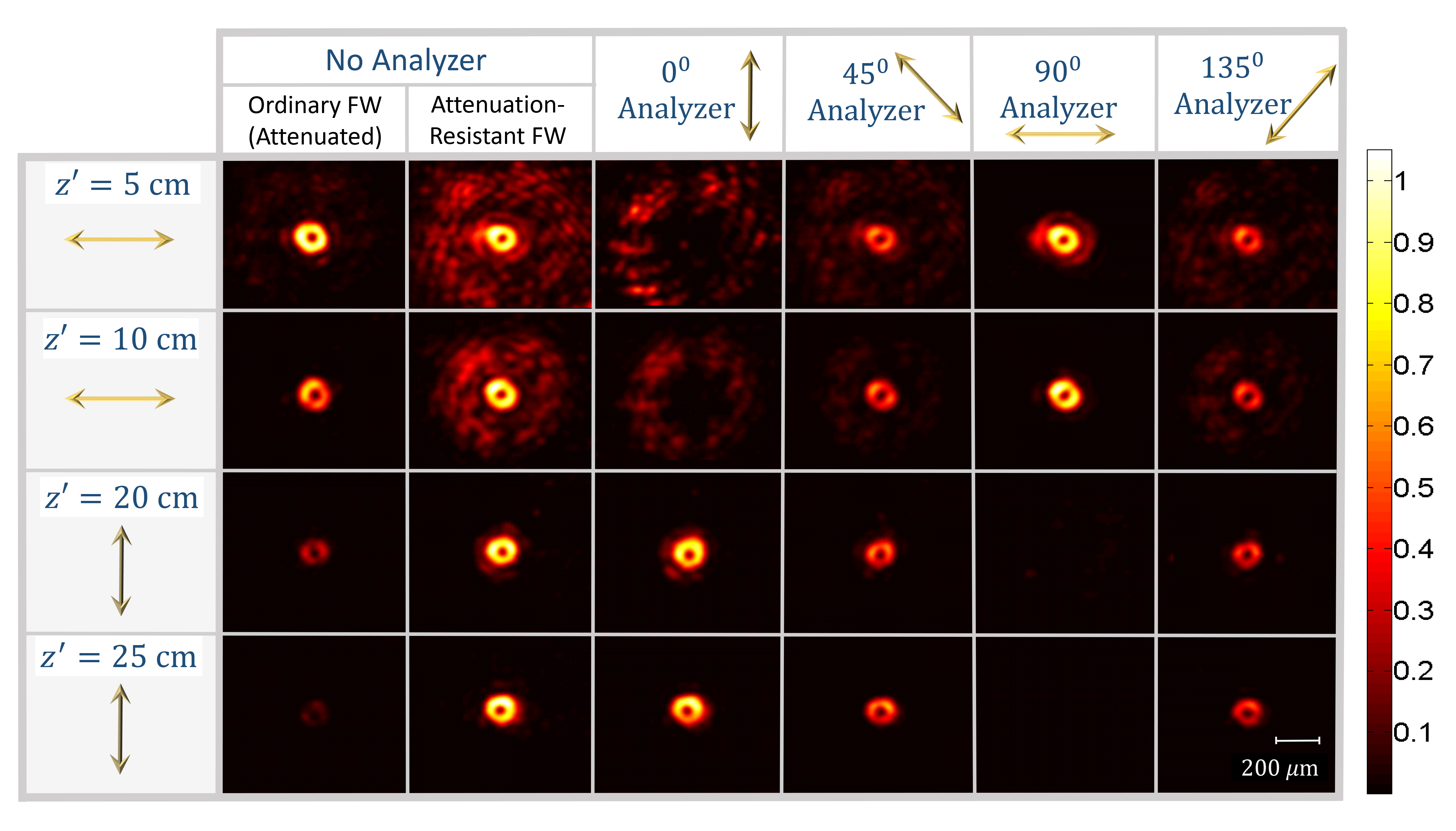}
	\caption{Normalized intensity profiles of the measured beam for the third example (generated in a lossy fluid). The top row indicates the analyzer angle used before the CCD camera to verify the change in SoP with propagation and the leftmost column indicates the expected SoP. The angles are measured with respect to the vertical axis $y$. All the columns not under the label ``No Analyzer'' refer to the attenuation-resistant beam. The beam is linearly polarized in the $x$-direction over the range $0\,\text{cm}\leq z^\prime < 15\,\text{cm}$. Then, it is made to posses a linear polarization in the $y$-direction with the same intensity level over the range $15\,\text{cm}\leq z^\prime < 30\,\text{cm}$, thus overcoming propagation losses. This can be better seen by comparing the two columns under the label ``No Analyzer'', the first of which presents the same beam without the loss compensation when choosing the complex amplitudes of the BBs.}
	\label{Fig7}
\end{figure*}

Here, $E_x$ and $E_y$ are engineered to maintain the same intensity level despite propagation losses. As mentioned in the footnote [37], attenuation-resistance is achieved by accounting for the medium losses when calculating the complex coefficients $A_{pm}$ via Eq.~\eqref{An}, which is done by appending the exponential term $e^{\beta_{i_0}z}$ to $F(z)$. The morphological functions are given by $G_{x1}=e^{i\phi}$, $G_{y1}=0$, $G_{x2}=0$, $G_{y2}=e^{i\phi}$ and $F_p(z)$ is defined as

\begin{align}
\begin{cases}
F_1(z)=1\,\text{,}  &F_2(z)=0 \qquad \! \underset{\text{\normalsize(x-polarization)}}{\text{for } 0\,\text{cm}\leq z^\prime < 15\,\text{cm}}
\\
F_1(z)=0\,\text{,}  &F_2(z)=1 \qquad \! \underset{\text{\normalsize(y-polarization)}}{\text{for } 15\,\text{cm}\leq z^\prime < 30\,\text{cm}}
\\
F_1(z)=0\,\text{,}  &F_2(z)=0 \qquad \! \text{elsewhere}
\end{cases}
\label{Equation4}
\end{align}

\par According to these parameters, the generated beam is $x$-polarized over the range $0\,\text{cm}\leq z^\prime < 15\,\text{cm}$, since it mainly has the contribution of $E_x$ in the beam center. Then, in the interval $15\,\text{cm}\leq z^\prime < 30\,\text{cm}$, the polarization is rotated such that the beam becomes $y$-polarized. This is done while maintaining the same intensity level along propagation inside the lossy fluid. The evolution of the transverse profile of the resulting beam recorded with the CCD camera at different propagation distances is shown in Fig.~\ref{Fig7}.

\par The transverse intensity pattern of the generated attenuation-resistant beam at different longitudinal positions is depicted in the second column under the label ``No Analyzer'' of Fig.~\ref{Fig7}. It is also contrasted with the intensity of a non-attenuation-resistant version of the FW superposition, that is, the same beam but without using the attenuation compensation technique when calculating the coefficients $A_{pm}$. In other words, they are calculated without the term $e^{\beta_{i_0}z}$ in Eq.~\eqref{An}. 

\par The evolution of the intensity pattern of this beam is shown in the first column under the label ``No Analyzer''. It is observed that its intensity decays exponentially with propagation due to medium absorption. This effect is clearly mitigated for the beam in the second column. Such attenuation-resistant property is a result of intensified self-healing process in which the outer rings of the beam act as an energy reservoir that intensify the central ring continuously with propagation---a result of the term $e^{\beta_{i_0}z}$ in Eq.~\eqref{An}. For example, in Fig.~\ref{Fig7}, the outer rings of the beam at $z^\prime=0\,\text{cm}$ for the attenuation-resistant beam are clearly pronounced when compared to the case of the ordinary beam. These rings are mainly $y-$polarized, as shown in the third column (under the ``$0^\circ$ Analyzer'' recordings). The energy in these rings is focused to intensify the central part of the beam throughout propagation. 
We note that such intensified self-healing behavior has been previously reported in \cite{FW_ab_3, absorption_1, absorption_2, absorption_3, absorption_4}. 

\begin{figure*}[htbp]
	\centering
	\includegraphics[width=.85\textwidth]{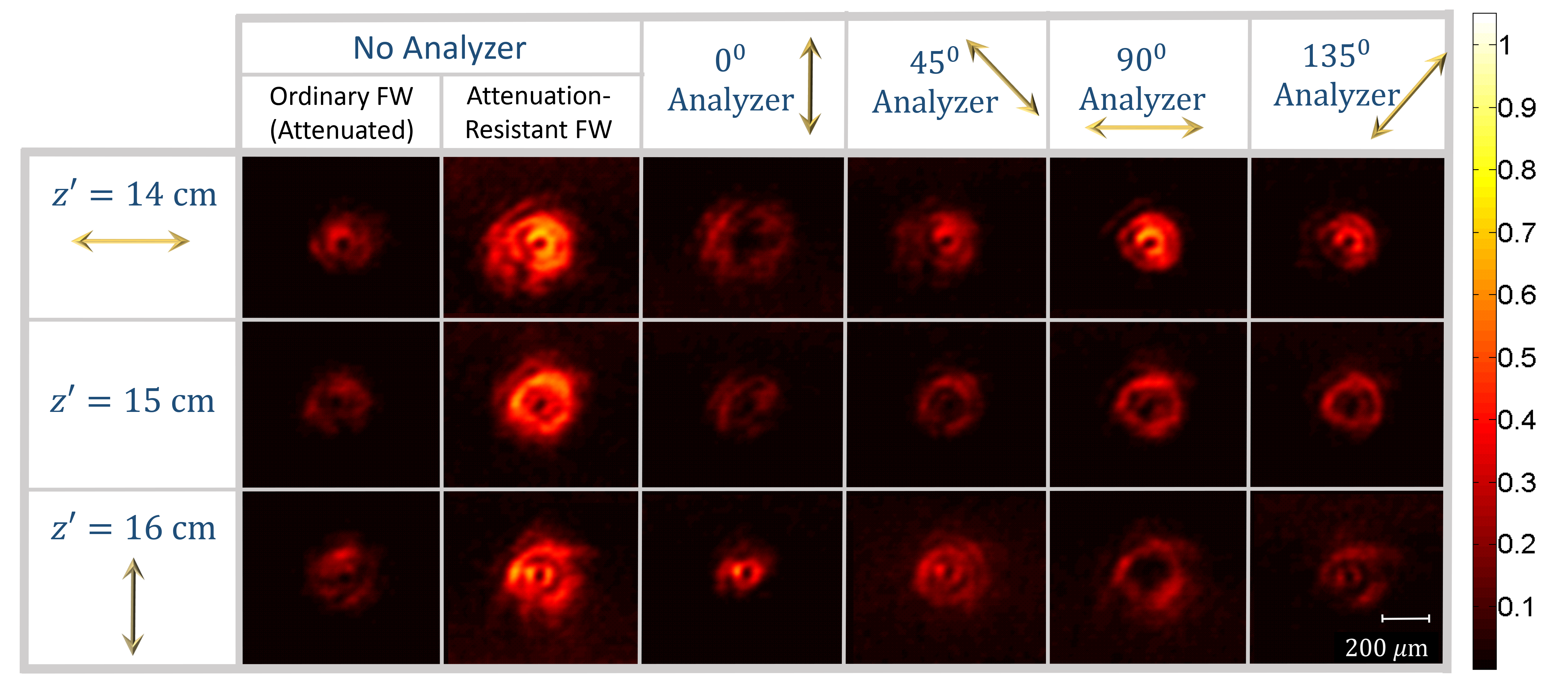}
	\caption{Normalized intensity profiles of the measured beam for the third example (generated in a lossy fluid) taken at three consecutive planes around the SoP transition: $z^\prime =14$, $15$, and $16\,\text{cm}$, for different analyzer angles. The figure depicts the transition of the beam from $x-$polarization to $y$-polarization.}
	\label{Fig8}
\end{figure*}

In order to investigate the transition of the beam from $x-$polarization to $y-$polarization, the beam evolution has been recorded around the $z^\prime = 15\,\text{cm}$ plane. The results are depicted in Fig.~\ref{Fig8}. It is observed that the $x-$polarized component of the beam washes away and diffuses to the outer rings with propagation, as observed in the column labeled ``$90^\circ $ Analyzer''. Meanwhile, the $y-$polarized component focuses and becomes more significant in the center, as observed in the column labeled ``$0^\circ$ Analyzer''. Since the available SLM only allows us to use 11 BBs in the superposition ($M = 5$), such transition was not very abrupt in comparison with the simulation results reported in Fig.~\ref{frames} (in which $M$ was set to 30). 

\par A few remarks are in order. First, using our proposed method, abrupt transitions in the SoP can be readily achieved by incorporating BBs with high spatial frequencies in the superposition, as theoretically shown in Sec.~\ref{example in lossy medium}. Experimentally, this requires using SLMs with small pixel pitch.  Second, the method we presented here to control the SoP and the intensity profile along propagation is compatible with the techniques of previous works on FWs that demonstrated control over other field characteristics (such as its transverse profile \cite{compound_FW} and topological charge \cite{FW_OAM}) and that showed propagation along off-axis trajectories \cite{spiral_snake}, thus allowing an additional simultaneous manipulation of other beam's properties. Finally, it is worth noting that FWs possess additional compelling features such as self-reconstruction and diffraction-resistance, which are inherent characteristics of Bessel beams.

\section{Conclusions}
\label{sec: conclusions}

\par We theoretically showed and experimentally demonstrated how a superposition of Frozen Waves (FWs) allows the arbitrary and simultaneous control of the polarization state and the intensity of the resulting beam as it propagates. Our technique provides a systematic and versatile method to engineer the longitudinal characteristics of a field in lossless and lossy media and, if combined with other FWs techniques, to also provide control over the beam's transverse structure. This can address many challenges in applications that benefit from the manipulation of diffraction-attenuation-resistant beams. In particular, our approach to control the state of polarization with propagation can be useful for material processing, polarimetry, microscopy and optical communications, to name a few.

\section{Acknowledgements}

\par This work was supported by FAPESP (grant 2015/26444-8), CNPq (grants 304718/2016-5 and 118966/2014-6) and the Natural Sciences and Engineering Research Council of Canada (NSERC - CGS program).

%\bibliographystyle{apsrev4-1}
%\bibliography{mybib}

\providecommand{\noopsort}[1]{}\providecommand{\singleletter}[1]{#1}%

\end{document}